\documentstyle[12pt]{article}
\begin{document}
\parskip 10pt plus 1pt
\title{Spacetime Dependent Lagrangians and the Vacuum Expectation Value of the
Higgs field}
\author{
{\it Rajsekhar Bhattacharyya}$^{*}$\\
{Dept.of Physics, Dinabandhu Andrews College, Kolkata-700084, INDIA}\\
{\it  Debashis Gangopadhyay$^{\dagger}$}\\ 
{S.N.Bose National Centre For Basic Sciences}\\
{JD Block, Sector-III, Salt Lake, Kolkata-700098, INDIA}\\
}
\date{}
\maketitle
\baselineskip=20pt
\begin{abstract}
The spacetime dependent lagrangian formalism of references
[1-2] is used to obtain a classical solution of Yang-Mills theory. This
is then used to obtain an estimate of the vacuum expectation value of the Higgs
field , {\it viz.} $\phi_{a}= A/e$, where $A$ is a constant and $e$ is the 
Yang-Mills coupling (related to the usual electric charge). The solution can
also  accommodate non-commuting coordinates on the boundary of the theory which
may be used to construct $D-$brane actions.

\end{abstract}
\newpage
The spacetimetime dependent lagrangian formalism [1-2] gives an alternative
way to deal with electromagnetic duality [3], weak-strong duality [4] and electro-
gravity duality [5]. Here this method will be used to obtain an estimate of the
vacuum expectation value of the Higgs field in terms of the electric charge
$e$ and a constant. The motivation of the present work comes from Ref.[2b] where
an analogue of the Bogomolny bound has been obtained for the Barriola-Vilenkin [6]
gravitational monopole. We will also show that the 't Hooft ansatz for
obtaining the "t Hooft-Polyakov monopole solution is sufficiently general to
lead to other solutions containing coordinates near the boundary that do 
not commute. Moreover, we will show that the 't Hooft ansatz for the gauge 
field is sufficient to yield a solution for the Higgs field for 
$r\rightarrow\infty$ without the necessity of any further ansatz for $\phi$.
We first briefly review the relevant material of [1].

Let the lagrangian $L'$  be a function of fields $\eta_{\rho}$, 
their derivatives $\eta_{\rho,\nu}$ {\it and the spacetime 
coordinates $x_{\nu}$}, i.e. $L'= L'(\eta_{\rho},\eta_{\rho,\nu}, x_{\nu})$.
Variational principle [12] yields :
$$\int dV \left(\partial_{\eta}L'
- \partial_{\mu}\partial_{\partial_{\mu}\eta}L'\right) = 0\eqno(1)$$
Assuming a separation of variables : 
$L'(\eta_{\sigma},\eta_{\sigma,\nu},.. x_{\nu})
=L(\eta_{\sigma},\eta_{\sigma,\nu})\Lambda(x_{\nu})$\\
($\Lambda(x_{\nu})$ is the $x_{\nu}$ dependent part and is a finite non-vanishing
function) gives
$$\int dV \left(\partial_{\eta}(L\Lambda)
- \partial_{\mu}\partial_{\partial_{\mu}\eta}(L\Lambda)\right) = 0\eqno(2)$$

We will be confined to classical solutions of theories where the fields do not
couple to gravity. Then $\Lambda$ is not  dynamical and is a finite,
non-vanishing function of  $x_{\nu}$ multiplying the primitive lagrangian $L$.
It is like an external field and equations of motion for $\Lambda$ meaningless. 
Duality invariance is related to finiteness of $\Lambda$. When equations of
motion are duality invariant, finiteness of $\Lambda$ on the spatial boundary
at infinity leads to new solutions for the fields. Poincare invariance and
duality invariance is achieved through same behaviour of $\Lambda$. 
The finite behaviour of $\Lambda$ on the boundary encodes the exotic
solutions of the theory within the boundary. In this way we are reminded of 
the holographic principle. 

Consider the Georgi-Glashow model with [3]
$$L= [-(1/4)G^{\mu\nu}_{a} G_{a\enskip\mu\nu} + (1/2)(D^{\mu}\phi)_{a}(D_{\mu}\phi)_{a} - V(\phi)]\eqno(3a)$$
where usually one takes $V(\phi)= (\lambda/4)(\phi^{a}\phi^{a}-a^{2})^{2}$.
The gauge group is $SO(3)$,$a,b,c$ are $SO(3)$ indices, with the generators
$\tau^{a}$ satisfying $[\tau^{a},\tau^{b}]=i\epsilon^{abc}\tau^{c}$.Gauge fields
$W_{\mu}=W_{\mu}^{a}\tau^{a}$ and the field strength is
$G^{\mu\nu}_{a}= \partial^{\mu}W^{\nu}_{a} 
- \partial^{\nu}W^{\mu}_{a} - e\epsilon_{abc}W^{\mu}_{b}W^{\nu}_{c}$,
$\tilde G^{\mu\nu}_{a}=(1/2)\epsilon^{\mu\nu\rho\sigma}G_{a\enskip\rho\sigma}$;
and the matter fields $\phi$ are in the adjoint representation of $SO(3)$.
The equations of motion are
$$(D^{\nu}G_{a\enskip\mu\nu})_{a}=\partial^{\mu}\phi_{a}-e\epsilon_{abc}W^{\mu}_{b}\phi_{c}\eqno(3b)$$
$$(D^{\mu}D_{\mu}\phi)_{a}=- \partial_{{\phi}^{a}}V\eqno(3c)$$
There are also the  Bianchi identities 
$$D^{\mu}\tilde G_{a\enskip\mu\nu}=0\eqno(3d)$$
Duality invariance means that $D^{\mu}G_{a\enskip\mu\nu}=0$.The energy density
is:
$$\Theta_{00}=(1/2)[(E^{i}_{a})^{2}+(B^{i}_{a})^{2}+(D^{0}\phi_{a})^{2}
+(D^{i}\phi_{a})^{2}+ V(\phi)]\eqno(4)$$
where the non abelian electric and magnetic fields are defined respectively as:
$E^{i}_{a}=-G^{0i}_{a}$  and  $B^{i}_{a}=-(1/2)\epsilon^{i}_{jk}G_{a}^{jk}$

The energy density $\Theta_{00}\geq0$ and the vacuum configuration is 
$$G^{\mu\nu}_{a}=0\enskip;\enskip D_{\mu}\phi=0\enskip;\enskip V(\phi)=0\eqno(5)$$

For this theory  $L'$ is then 
$$L'= L\Lambda=[-(1/4)G^{\mu\nu}_{a} G_{a\enskip\mu\nu} 
+ (1/2)(D^{\mu}\phi)_{a}(D_{\mu}\phi)_{a} 
- V(\phi)]\Lambda(x_{\nu})\eqno(6)$$
Equations of motion using $(2)$ are :
$$\Lambda (D^{\mu}G_{a\enskip\mu\nu}) + (\partial^{\mu}\Lambda)G_{a\enskip\mu\nu}
+ \Lambda e \epsilon_{abc} (\partial_{\nu}\phi)_{b}(\phi)_{c} 
- \Lambda e^{2} \epsilon_{abc}\epsilon_{bc'd'}W_{\nu\enskip c'} 
\phi_{c} \phi_{d'} = 0\eqno(7a)$$   
$$(D^{\mu}D_{\mu}\phi)_{a}\Lambda + (D_{\mu}\phi)_{a}\partial_{\mu}\Lambda 
= - (\partial_{{\phi}^{a}}V)\Lambda\eqno(7b)$$
and the Bianchi identities are:
$$D^{\mu}\tilde G_{a\enskip\mu\nu}=0\eqno(7c)$$
Requiring duality invariance (i.e.$D^{\mu}G_{a\enskip\mu\nu}=0$) gives 
$$(\partial^{\mu}\Lambda)G_{a\enskip\mu\nu} 
= - \Lambda e \epsilon_{abc} (D_{\nu}\phi)_{b}(\phi)_{c}\eqno(8)$$ 
For $\Lambda=\Lambda(r)$ we have 
$$\Lambda_{\infty} 
= \Lambda_{0} exp[-e \int_{0}^{\infty} dr \left( (\epsilon_{abc} 
(D_{\nu}\phi)_{b} \phi_{c}) 
(\partial^{i} r G_{a \enskip i\nu})^{-1}\right)]\eqno(9)$$
where $\Lambda_{p}$ is the value of $\Lambda$ at $r=p$;
$a,\nu$ are fixed; and there is a sum over indices $i,b$ and $c$.
$\Lambda_{\infty}$ must be finite. Choose this
to be the constant unity.This may be realised in various ways, the simplest being
$(D_{\nu} (\phi)_{b} \Rightarrow 0$, \enskip 
$(\phi)_{c} \Rightarrow finite$, \enskip
and the product
$(D_{\nu}\phi)_{b} (\phi)_{c}$ falls off faster than $G_{a \enskip i\nu}$ for
large $r$. Then a constant value for $\Lambda$ is perfectly consistent with $(7b)$ and
the conditions become analogous to the Higgs' vacuum condition for
the t'Hooft-Polyakov monopole solutions where the duality invariance of the 
equations of motion and Bianchi identities are attained at large $r$ by 
demanding $(D_{\mu} \phi)_{a} \Rightarrow 0$ and $\phi_{a}\Rightarrow a\delta_{a3}$
at large $r$. Note that our results are perfectly consistent with the 
usual choice for the Higgs' potential $V(\phi)$ even though nothing has been 
assumed regarding this. Thus, the t'Hooft-Polyakov monopole solutions 
follow naturally in our formalism. 
We now discuss two other interesting possibilities.

{\bf Case I}

$$\left(\epsilon_{abc} (D_{\nu}\phi)_{b}\phi_{c}\right) \Rightarrow 0\eqno(10)$$ 
(i.e. the duality condition $(D^{\mu}G_{\nu\mu})_{a}=0$)
and falls off faster than $G_{a \enskip i\nu}$ for large $r$ 
($a$ and $\nu$ are fixed). A solution is when 
$$D_{\nu}  \phi = \alpha_{\nu} \phi\eqno(11)$$
where $\alpha_{\nu}$ can be any Lorentz four vector field that is consistent with all 
the relevant equations of motion and the minimum
energy requirements. The minimum energy requirements are satisfied because it
is straightforward to verify that the gauge fields $W^{\mu}_{a}$ do not change.
This is seen by taking
the cross product of $\phi$ ($\phi$ is a $SO(3)$ vector)
with equation $(11)$. We again arrive at the well known results of Corrigan 
{\it et al} [7], {\it viz.} $W^{\mu}= (1/a^{2}e)\phi\wedge\partial^{\mu}\phi +(1/a)\phi A^{\mu}$,
where $A^{\mu}$ is arbitrary.

As the gauge fields $W^{\mu}_{a}$ do not change, so even with this solution 
we obtain the same gauge field solutions as before and so minimum energy 
requiremet is automatically satisfied. However, this new solution allows us 
to obtain an estimate of the vacuum expectation value of the Higgs field and
to this we now proceed. Let 
$\alpha_{\nu}=(0,\alpha_{i})\equiv \alpha(r) \hat r$, where $\hat r$ is the unit
radial vector.So the Bogomolny condition is $B^{i}_{a}=D^{i}\phi_{a}=\alpha^{i}\phi_{a}$ 
and the Higgs vacuum condition obtained from equation $(7b)$ 
(for $r\rightarrow \infty, \Lambda$ is a constant , say unity) is 
$$[D^{i}(\alpha_{i}\phi)]_{a}=-\partial_{\phi^{a}} V = 0\eqno(12a)$$
i.e. we are at a minima of the potential $V$. If $\phi_{a}\not= 0$, $(12a)$ implies
$$div\enskip \vec{\alpha} + \vec{\alpha}^{2} =0\eqno(12b)$$
and the solution is 
$$\alpha(r)= 1/(cr^{2}-r)\enskip , \alpha^{i}=r^{i}/(cr^{3}-r^{2})\eqno(13)$$
where we take the constant $c$ to be negative.
Let us now take the 't Hooft ansatz for the gauge field, {\it viz.}
$$W^{0}_{a}=0\enskip \enskip ;  W^{i}_{a}= -\epsilon_{aik}r^{k}[1-K(aer)]/(er^{2})\eqno(14)$$
where the function $K(aer)$ has been well studied [3] and goes to zero at $r\rightarrow\infty$.
Then  the electric field vanishes while $G_{a\enskip jk}, B^{i}_{a}$ are
$$G_{a\enskip jk}
=(1/er^{2})[2\epsilon_{ajk}(1-K)
+\epsilon_{akl}r^{l}\partial_{j}K-\epsilon_{ajl}r^{l}\partial_{k}K]$$
$$+(1/er^{4})[2(1-K)(\epsilon_{akl}r^{l}r_{j}-\epsilon_{ajl}r^{l}r_{k})
+(1-K)^{2}(\delta_{aj}\epsilon_{ckl}r^{c}r^{l}-\epsilon_{jkl}r_{a}r^{l})]\eqno(15a)$$
$$B^{i}_{a}= (1/er^{4})[(1-K)^{2}r_{a}r^{i}-2(1-K)r^{i}r_{a}]
-(1/er^{2})[r^{i}\partial_{a}K-\delta^{i}_{a}r^{m}\partial_{m}K]\eqno(15b)$$
Now $B^{i}_{a}=D^{i}\phi_{a}=\alpha^{i}\phi_{a}$.Therefore,taking $c=-A$ so that $A$
is positive we have
$$\phi_{a}={(1+Ar)\over{(er^{2})}}[2(1-K)r_{a}-(1-K)^{2}r_{a}+r^{2}\partial_{a}K-r_{a}r^{m}\partial_{m}K]\eqno(15c)$$
It is easily seen that $(15c)$ reduces to the 't Hooft ansatz for $\phi_{a}$
for $A=0$ and $r\rightarrow\infty$. Thus we have obtained an expression for
$\phi$ without assuming any ansatz. This had never been possible before.
There is another interesting outcome.For $r\rightarrow\infty$ we have
$K\rightarrow 0$ and so 
$$\phi_{a}\rightarrow {Ar^{a}\over er} + {r^{a}\over er^{2}}
={A\over {e}}\hat r^{a} +{\hat r\over {er}}
\rightarrow{A\over {e}} \hat r^{a}\eqno(16)$$
for $r\rightarrow\infty$. But 
$\phi_{a}=a\delta_{a3}$ for $r\rightarrow\infty$. Therefore $a=A/e$.
This is the principal result of this work.

{\bf Case II}

$\alpha_{\nu}$ is any Lorentz four vector field as in {\bf I} but which 
may also carry internal symmetry indices {\it other than $SO(3)$}
with the generators of the symmetry satisfying some Lie algebra 
$[T_{P},T_{Q}]= if_{PQR} T_{R}$. Let us take the group to be $SU(2)$.
i.e. say, $\alpha^{\nu}= \alpha^{\nu}_{P}T_{P}$; $P,Q,R=1,2,3$;
$T_{P}$ being the generators of $SU(2)$. 
Again choosing $\alpha^{\nu}=(0,\alpha^{i}_{P}T^{P})$ with
$\vec\alpha_{P}=\alpha_{P}(r)\hat r$ and
using the well known properties of 
the Pauli matrices it is easily seen that  the analogue of equation $(12)$ is
$$div\enskip \vec{\alpha_{P}}=0\eqno(17)$$
which has the solution
$$\alpha^{i}_{P}=A_{P} {r^{i}\over {r^{3}}}\eqno(18)$$
where $A_{P}$ are constants. Writing $r^{i}_{P}=A_{P} r^{i}=r^{3}\alpha^{i}_{P}$,
we can then define new coordinates 
$$R^{i}=r^{i}_{P}T_{P}\enskip ; [R^{i},R^{j}]\not=0\eqno(19)$$
and these are non-commuting. Moreover, they carry both Lorentz and internal
indices and hence are like gauge fields in some different theory. Note that 
transverse coordinates (i.e. transverse to the brane and lying in the bulk volume)
in $D$ brane theories are often identified with gauge
fields [8] and so we can construct such actions with our solutions $(19)$.
Under these circumstances, equation  $(11)$ should be written as 
$$\partial_{\mu}^{P}\phi_{a}-e\epsilon_{abc}(W_{\mu}^{b})^{P}\phi_{c}=\alpha_{\mu}^{P}\phi_{a}$$
where the coordinates and their differentials are now matrices and capital
alphabets denote the indices of the new symmetry group. For fixed $P$,  
$(W_{\mu}^{b})^{P}$ may be identified with the old gauge fields $W_{\mu}^{b}$.

A point to note is that we have taken the symmetry group for $\alpha_{\nu}$ to be 
some group other than $SO(3)$. This is to ensure in the simplest possible way 
that the fields $(W_{\mu}^{b})^{P}$ {\it for fixed P} may be identified with the
old (i.e.unchanged) gauge fields $W_{\mu}^{b}$ ($P$ is now a fixed index) and 
so the minimum energy requirements are satisfied in each sector of $P=1,2,3$.
(This is seen by taking the cross product of $\phi$ with the analogue of 
equation $(11)$ and proceeding as before).So each sector now contains a monopole.
Then we have a configuration that is quite similar to "string" of monopole 
solutions connecting two D-branes. Such configurations are known in the literature [8].
The other point is that the vacuum expectation value of the Higgs field is 
proportional to the inverse of the coupling $e$; and this result has been 
obtained from the classical solutions. This result is similar to that obtained
in Ref.[2b] if we are ready to identify the inverse of Newton's gravitational
constant (which is definitely the coupling constant in theories of gravity)
as the vacuum expectation value of some field hitherto unknown.

The solutions in equation $(11)$ were hidden in 't Hooft-Polyakov's 
work. This had been overlooked before for the simple reason because at that point
of time one was more concerned in obtaining solutions from the minimum 
(finite ) energy principles. This was perfectly justified. We have obtained
the solutions  from the requirement of duality invariance which is quite relevant
at this point of time. However, we have also shown that the duality requirements
automatically contain the minimum energy condition ($\Lambda$ is also finite for $D_{\nu}\phi=0$).

All the results have been obtained at $r\rightarrow\infty$. That is, we are at the boundary 
of the theory. So the finiteness of $\Lambda$ at the boundary encodes the duality
invariance of the theory within the boundary and thus an analogue of the 
holographic principle [9] seems to be at work. On the boundary there seems to
exist {\it a different gauge field theory} together with non-commuting coordinates.

In conclusion, the spacetime dependent lagrangian formalism in conjunction
with the 't Hooft-Polyakov results have yielded an expression for the 
vacuum expectation value of the Higgs field as $A/e$. This result is definitely
susceptible to experiments. We have also shown that the 't Hooft ansatz for the
gauge field is sufficient to obtain an expression for the Higgs field if one
uses our formalism. No additional ansatz for $\phi$ is necessary.
The expression obtained reduces to the 't Hooft ansatz for
the Higgs field at $r\rightarrow\infty$. Finally, we have shown that classical solutions
of Yang-Mills theory also contain the germ of non-commuting coordinates 
residing on the boundary. The structure of these coordinates are like 
gauge fields and hence are relevant in constructing $D$-brane actions.

$^{*}$Electronic address : rajsekhar@vsnl.net

$^{\dagger}$Electronic address : debashis@boson.bose.res.in

\end{document}